\documentclass[aps,twocolumn,showpacs,floatfix]{revtex4}

\usepackage{amsmath}

\usepackage{graphicx}

\begin{document}

\title{Bloch-like oscillations in a one-dimensional lattice with long-range
correlated disorder}

\author{F.\ Dom\'{\i}nguez-Adame}
\author{V.\ A.\ Malyshev}
\thanks{On leave from ``S.I. Vavilov State Optical Institute'', Birzhevaya
Linia 12, 199034 Saint-Petersburg, Russia}
\affiliation{GISC, Departamento de F\'{\i}sica de Materiales, Universidad
Complutense, E-28040 Madrid, Spain}

\author{F.\ A.\ B.\ F.\ de Moura}
\author{M.\ L.\ Lyra}
\affiliation{Departamento de F\'{\i}sica, Universidade Federal de Alagoas,
Macei\'{o} AL 57072-970, Brazil}

\begin{abstract}

We study the dynamics of an electron subjected to a uniform electric field
within a tight-binding model with long-range-correlated diagonal disorder.
The random distribution of site energies is assumed to have a power
spectrum $S(k) \sim 1/k^{\alpha}$ with $\alpha > 0$. Moura and Lyra [Phys.\
Rev.\ Lett.\ {\bf 81}, 3735 (1998)] predicted that this model supports a
phase of delocalized states at the band center, separated from localized
states by two mobility edges, provided $\alpha > 2$. We find clear
signatures of Bloch-like oscillations of an initial Gaussian wave packet
between the two mobility edges and determine the bandwidth of extended
states, in perfect agreement with the zero-field prediction.

\end{abstract}


\maketitle


The single-parameter scaling hypothesis predicts localization of a single
quasiparticle in one (1D) and two dimensions with time-reversal symmetry,
independently of the disorder strength present in the
system~\cite{Abrahams79}. There exist, however, low-dimensional systems
that do not obey the single-parameter scaling framework. Thus, the absence
of Anderson localization in the presence of spatial short-range
correlations in disorder~\cite{Flores89,Dunlap90} was put forward to
explain transport properties of semiconductor superlattices with
intentional correlated disorder~\cite{Bellani99}. Further, it was
demonstrated that long-range correlated diagonal~\cite{Moura98,Izrailev99}
and off-diagonal~\cite{Moura99} disorder also acts towards delocalization
of 1D quasiparticle states. Furthermore, long-range correlations can result
in the emergence of a phase of extended states in the thermodynamic limit.
This phase appears at the band center and is separated from localized
states by two mobility edges~\cite{Moura98}. This theoretical prediction
was experimentally validated by measuring microwave transmission spectra of
a single-mode waveguide with inserted correlated scatterers~\cite{Kuhl00}.
In the case of short-range correlated disorder the delocalized phase does
not appear: the number of delocalized states increases proportionally to
the square root of the system size, and thus this phase has zero measure in
the thermodynamic limit.

In this Letter we focus on the dynamical properties of an electron in a
system with long-range correlated diagonal disorder. Interplay between the
delocalization effect, preserved by the long-range correlated disorder, and
the dynamic localization, caused by an electric field acting on the system,
is of our interest. We compute the behavior of an initial Gaussian wave
packet in the presence of a uniform electric field solving numerically the
1D time-dependent Schr\"{o}dinger equation for the complete Hamiltonian. We
found clear signatures of Bloch-like oscillations~\cite{Bloch28} of the
wave packet between the two mobility edges of the delocalized phase of
states. The amplitude of the oscillatory motion of the centroid allows us
to determine the bandwidth of the delocalized phase.


We consider a tight-binding Hamiltonian with long-range-correlated diagonal
disorder and an external dc electric field on a regular 1D open lattice of
spacing  $a$~\cite{Dunlap86}
\begin{align}
{\cal H} & = \sum_{n=1}^{N}\Big(\widetilde{\varepsilon}_{n}-e{\cal F}an\Big)
|n\rangle\langle n| \nonumber \\
 & -J\sum_{n=1}^{N-1}\Big(|n\rangle\langle n+1|+|n+1\rangle\langle n|\Big)\ ,
\label{hamiltonian}
\end{align}
where $|n\rangle$ is a Wannier state localized at site $n$ with energy
$\widetilde{\varepsilon}_{n}$, $\cal F$ is the external uniform electric
field, and $N$ is even. The intersite coupling is restricted to
nearest-neighbors and assumed to be uniform over the entire lattice with
$J>0$. In terms of the Wannier amplitudes $\psi_{n}(t)= \langle
n|\Psi(t)\rangle$, the Schr\"{o}dinger equation reads~\cite{Nazareno99}
\begin{equation}
i\dot{\psi}_{n}=(\varepsilon_{n}-Fn)\psi_{n}-\psi_{n+1}-\psi_{n-1}\ ,
\label{Schrodinger}
\end{equation}
where we introduced the dimensionless parameters $\varepsilon_{n} =
\widetilde{\varepsilon}_{n}/J$, $F = e{\cal F}a/J$, and time is expressed
in units of $\hbar/J$.

The source of disorder is the stochastic fluctuations of energies
$\epsilon_{n}$, which we are going to consider as being long-range
correlated. One of the simplest ways to numerically generate a power-law
correlated sequence of on-site potentials $\varepsilon_{n}$ is to write its
Fourier decomposition as follows~\cite{Moura98}
\begin{equation}
\varepsilon_{n} = C(\alpha)\sum_{k=1}^{N/2}\frac{1}{k^{\alpha/2}}\,
\cos\Bigg(\frac{2\pi nk}{N}+\phi_k\Bigg)\ .
\label{disorder}
\end{equation}
Here, $\phi_k$ are $N/2$ independent random phases uniformly distributed
within the interval $[0,2\pi]$, and $C(\alpha)$ is a normalization
constant. We will normalize the energy sequence to have $\langle
\varepsilon_{n}\rangle = 0$ and $\langle \varepsilon_{n}^2\rangle=1$, where
$\langle \ldots \rangle$ indicates average over the random phases
$\phi_k$.  The long-range nature of the potential correlations results from
the power-law dependence of the amplitudes on the wave-vector
characterizing each Fourier component. Several stochastic processes in
nature are known to generate long-range correlated random sequences which
have no characteristic scale, for example, in the nucleotide sequence of
DNA molecules~\cite{Peng92}. The relevance of the underlying long-range
correlations for the electronic transport in DNA has been recently
discussed in Ref.~\cite{Carpena02}. Furthermore, interface roughness
appearing during growth often displays height-height correlations with
power-law spectra~\cite{Barabasi95}; thus, the subsequent random potential
arising from the rough interface would be long-range correlated. Recently,
transport properties of systems with long-range correlated disorder was
explored, both theoretically and experimentally, in the design of devices
for filtering of electrical and optical signals~\cite{Krokhin02}.

Having introduced the model of disorder, we can solve
numerically~(\ref{Schrodinger}) by means of an implicit integration
algorithm~\cite{Press86}. The initial condition is set to a Gaussian wave
packet of width $\sigma$ and centered at $n_0=N/2$
\begin{equation}
\psi_{n}(0)=A(\sigma)\,\exp\left[-(n-n_0)2/4\sigma2\right]\ ,
\label{Gaussian}
\end{equation}
$A(\sigma)$ being the normalization factor. Once Eq.~(\ref{Schrodinger}) is
solved for the initial condition~(\ref{Gaussian}), we compute the mean position
of the wave packet (centroid)
\begin{equation}
x(t)= \sum_{n=1}^{N}(n-n_0)\,|\psi_{n}(t)|^2\ .
\label{tools}
\end{equation}
It is to be noticed that all eigenstates of the
Hamiltonian~(\ref{hamiltonian}) contribute to $x(t)$.


It is well known that a uniform electric field applied to a periodic
lattice causes the dynamic localization of the initially extended states of
an electron and gives rise to an oscillatory behavior of the electron wave
packet, the so-called Bloch oscillations~\cite{Bloch28}. The length of the
segment over which the electron oscillates (twice the amplitude) and the
period of these oscillations are estimated from semiclassical arguments to
be (in dimensionless units) $L_F = W/F$ and $\tau_B=2\pi/F$ respectively
(e.g., see Ref.~\cite{Ashcroft76}), where $W$ is the width of the Bloch
band in units of the coupling integral $J$ ($W = 4$ in our case). Notice
that this approach requires slowly varying wave functions to be valid,
implying that $L_F > 1$. Similar condition was pointed out in
Ref.~\cite{Arrachea02}. The question we aim to clarify is to what extent
will the corresponding phenomenology be valid if long-range correlated
disorder, that allows for a delocalized phase, is present. In other words,
whether the biased phase of extended states bounded by two mobility edges
behaves similarly to a biased Bloch band. If so, it provides us with a
method to measure the energy width of the delocalized phase $W^\prime$ from
the relationship $L_F^\prime = W^\prime/F$,  where $L_F^\prime$ is twice
the amplitude of the oscillatory motion of the centroid. Below,  we present
a numerical proof of this conjecture.

\begin{figure}[ht]
\centerline{\includegraphics[width=56mm,clip]{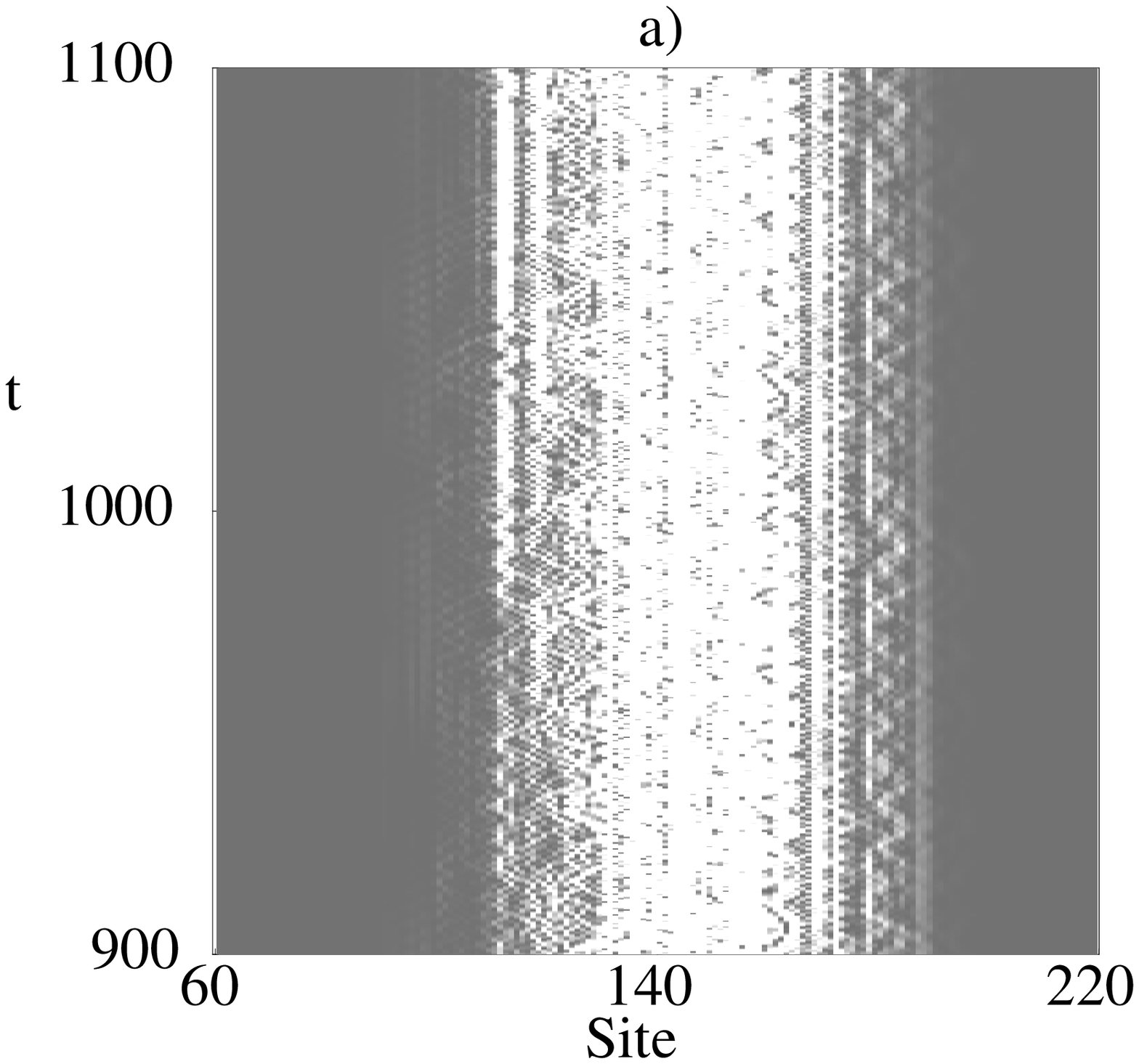}}
\centerline{\includegraphics[width=56mm,clip]{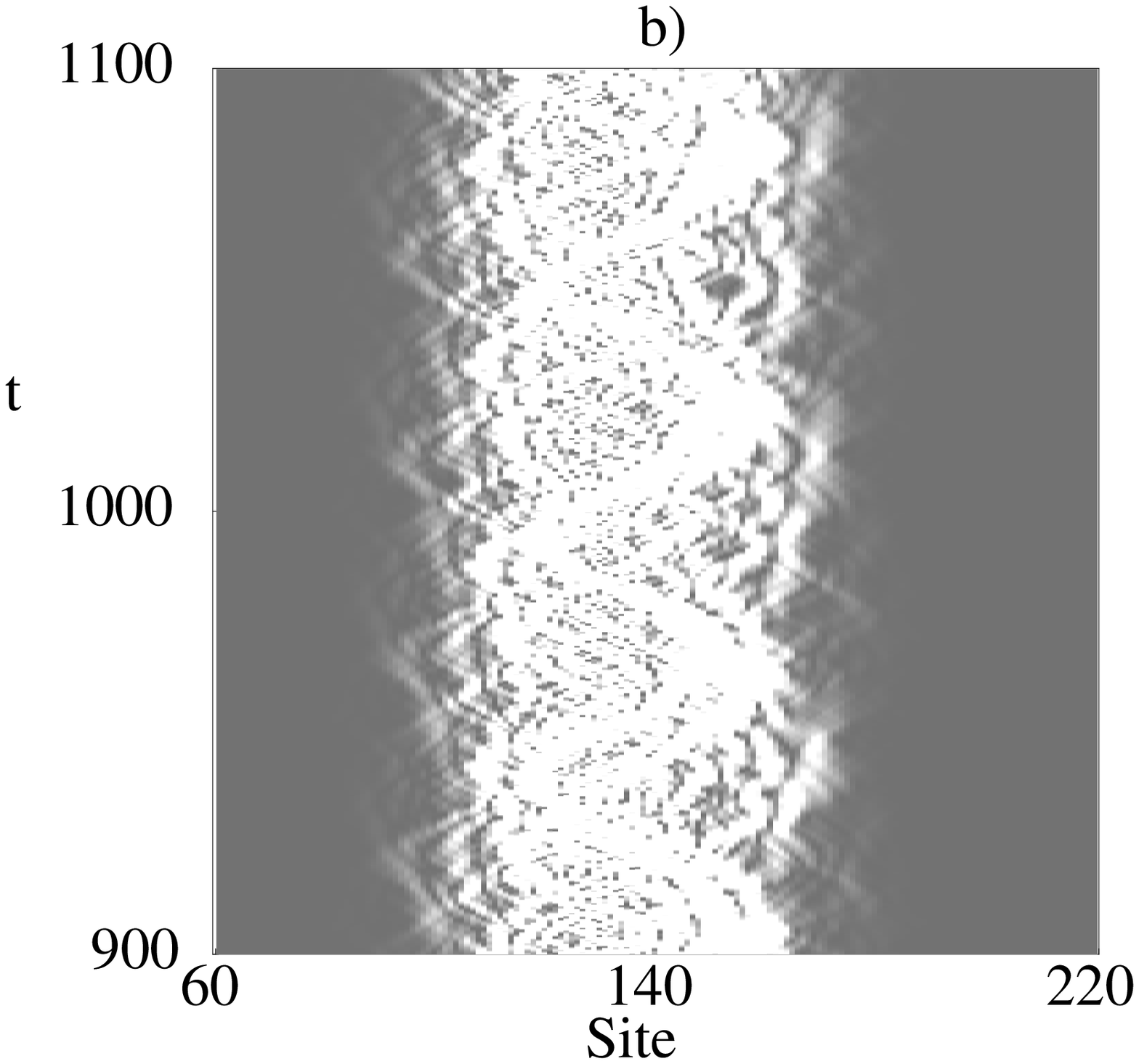}}
\caption{Asymptotic dynamics ($t\gg\tau_B$) of the Wannier amplitudes
$\psi_{n}(t)$ of a biased wave packet ($\sigma = 10$ at $t = 0$ and $F =
0.1$) in a lattice with $N=300$ sites, for a) $\alpha=0.5$ and b)
$\alpha=3.0$. Bright and dark regions indicate higher and lower probability
amplitudes, respectively.}
\label{fig1}
\end{figure}

In Fig.~\ref{fig1} we show the asymptotic dynamics ($t\gg\tau_B$) of the
Wannier amplitudes $\psi_{n}(t)$ of a initial Gaussian wave packet
$\psi_n(t)$ ($\sigma = 10$ at $t = 0$ and $F = 0.1$) calculated for two
values of the exponent $\alpha$, 0.5 (the phase of delocalized states is
absent in the unbiased system) and 3.0 (the phase of  delocalized states is
present in the unbiased system). At $\alpha=0.5$ [see  Fig.~\ref{fig1}a)],
the disorder is almost uncorrelated and the system is similar  to the
standard Anderson model, with no signatures of Bloch oscillations. On  the
contrary, the plot in Fig.~\ref{fig1}b) calculated for $\alpha=3.0$ shows
an oscillating in time pattern. This result suggests that Bloch
oscillations can take place \emph{even in the presence of disorder}.

\begin{figure}[ht]
\centerline{\includegraphics[width=60mm,clip]{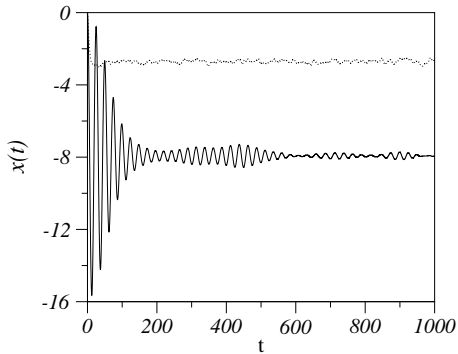}}
\caption{Time-domain dynamics of the centroid of a biased wave packet
($\sigma=10$ at $t = 0$ and $F = 0.25$) in a lattice  with $N=500$ sites.
Dotted lines correspond to $\alpha=0.5$ and  solid lines to $\alpha=3.0$.}
\label{fig2}
\end{figure}

The time-domain evolution of the centroid, $x(t)$, provides more detailed
information about what is happening. In Fig.~\ref{fig2} one can see that
there are no signatures of Bloch oscillations for $\alpha = 0.5$ (dotted
line), at least for a moderate field amplitude. Oscillations, which are
present at the very beginning, achieve in a short time a weakly fluctuating
(stationary  in average) value. This is a clear indication of the existence
of disorder-induced decoherence effects, as is expected for the localized
regime. On the contrary, at $\alpha=3.0$ the centroid displays an
oscillatory, amplitude-modulated pattern after an initial transient, where
the amplitude of the oscillations is reduced (solid line). This oscillatory
displacement of the wave packet is not accompanied by in-phase oscillation
of its width. In other words, breather modes are absent for the specified
initial condition.

\begin{figure}[ht]
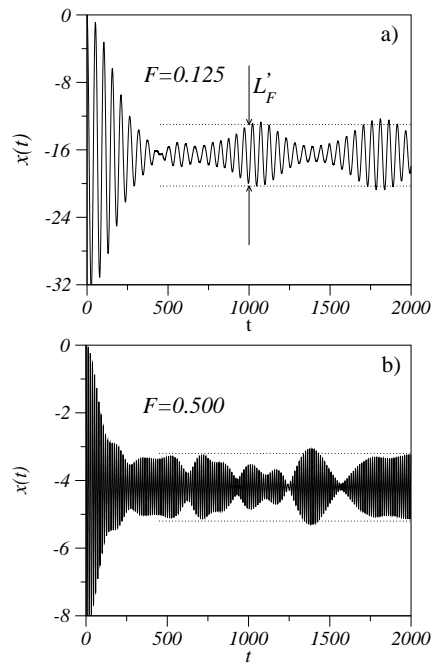

\centerline{\includegraphics[width=56mm,clip]{figure3a.eps}}
\centerline{\includegraphics[width=56mm,clip]{figure3b.eps}}
\caption{Time-domain dynamics of the centroid of a biased wave packet
($\sigma=10$ at $t = 0$) in a lattice with $N=1000$ when $\alpha=3.0$, for
a) $F=0.125$ and b) $F=0.5$. Dotted lines bound a region of size
$L_F^\prime$ within which the wave packet oscillates for a long time.}
\label{fig3}
\end{figure}

Valuable information can be extracted from a detailed inspection of the
centroid dynamics. First, we notice that the period of the oscillations is
a well defined quantity, as is seen from Fig.~\ref{fig3}. Furthermore,
within the numerical uncertainty, its value is equal to $\tau_B = 2\pi/F$,
being the period of Bloch oscillations in a homogeneous lattice. At the
initial stage of the motion, the spatial region within which the wave
packet oscillates is roughly given by $4/F$, as for an ideal Bloch band.
Disorder results in a relatively fast decrease of the oscillation
amplitude. After this transient stage, oscillations are not damped further
and remains amplitude-modulated with an envelope which depends on the
realization of disorder. Dotted lines in Fig.~\ref{fig3} bound the spatial
region within which the wave packet oscillates for a long time. On average,
the width of this region $L_F^\prime$ is found to be $L_F^\prime \sim
W^\prime/F$, where $W^\prime$ is independent of the applied field $F$. From
the data  in Fig.~\ref{fig3} we obtain $W^\prime \sim 1$. This value agrees
remarkably well with the width of the band of extended states reported in
Ref.~\cite{Moura98}  for $\alpha=3.0$. Thus, we arrive at the  main
conclusion of this work, namely  there exist clear signatures of Bloch-like
oscillations of a biased Gaussian  wave packet between the two mobility
edges.

\begin{figure}
\centerline{\includegraphics[width=60mm,clip]{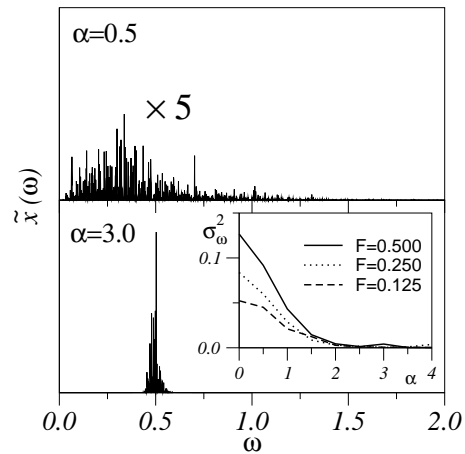}}
\caption{Fourier transform of the centroid of a biased wave packet
($\sigma=10$ at $t = 0$ and $F = 0.5$) in a lattice with $N=500$ sites,
for  $\alpha=0.5$ (upper panel and notice the magnification factor) and
$\alpha=3.0$ (lower panel). The inset shows the variance
$\sigma^{2}_{\omega}$ of $\widetilde{x}(\omega)$ as a function of $\alpha$
for different values of the applied electric field. Results were obtained
by averaging over $1000$ realizations of the disorder.}
\label{fig4}
\end{figure}

To provide further confirmation of this claim, we calculated numerically
the Fourier transform of the centroid, $\widetilde{x}(\omega)$, as shown in
Fig.~\ref{fig4}. Results were obtained by averaging over 1000 realization
of disorder. For $\alpha = 0.5$, the Fourier transform
$\widetilde{x}(\omega)$ is rather broad, suggesting that $x(t)$ is similar
to a white noise signal. On the contrary, for $\alpha =  3.0$  the Fourier
transform $\widetilde{x}(\omega)$ shows a well-defined, narrow peak at
about $\omega = F$, despite averaging.  The variance of
$\widetilde{x}(\omega)$, $\sigma^{2}_{\omega}$, displays a rather different
trend below and above $\alpha=2.0$, as seen in the inset of
Fig.~\ref{fig4}. For $\alpha <2.0$ the Fourier transform is broad, but for
$\alpha > 2.0$ the width of the Fourier transform is rather small and
independent of the electric field. The onset for the appearance of Bloch
oscillations $\alpha=2$ (see the inset of Fig.~\ref{fig4}) is in excellent
agreement with the value obtained in Ref.~\cite{Moura98} for the occurrence
of the phase of extended states.

\begin{figure}
\centerline{\includegraphics[width=60mm,clip]{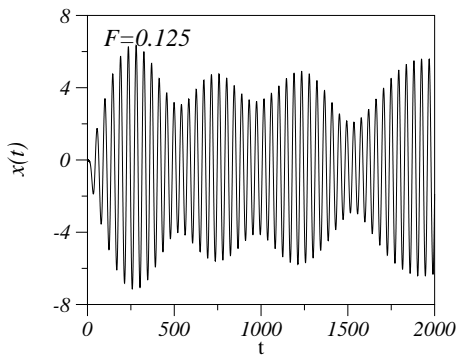}}
\caption{Time-domain dynamics of the centroid of a biased wave packet, with
initial condition $\psi_n(0)=\delta_{n\,n_0}$, in a lattice with $N=1000$
sites when $\alpha=3.0$, for $F=0.125$}
\label{fig5}
\end{figure}

Before concluding, some words concerning the influence of the width of the
initial wave packet are in order. The above results for $F\neq 0$ were
obtained with a relatively broad wave packet ($\sigma \gg 1$) with initial
velocity equal to zero. Consequently, its Fourier components span a very
narrow set of states at the bottom of the band. We performed also a study
for  initial moving wave packets located initially deep in the band. In
short, in the very beginning the wave packet explores the entire band in
either case and oscillates within a spatial region of size $4/F$. On
increasing time, the amplitude of $x(t)$ is always reduced to a region of
size $L_F^\prime = W^\prime/F$, where $W^\prime$ is the width of the
delocalized phase. This picture changes slightly for narrower initial wave
packets, for which the large-amplitude transient is absent.
Figure~\ref{fig5} shows the centroid dynamics of an initial Kronecker
$\delta$ wave packet at $\alpha=3.0$ and $F=0.125$. The extent of the
spatial region within which the wave packet asymptotically oscillates is
again $L_F^\prime = W^\prime/F$. Thus, we can confidently state that
Bloch-like oscillations are rather insensitive to initial conditions.

We studied a biased random tight-binding model where the on-site disorder
is long-range correlated  with the power spectrum $S(k) \sim 1/k^{\alpha}$,
\, $\alpha > 0$. The unbiased model supports a phase of delocalized states
in the center of the band provided $\alpha > 2$~\cite{Moura98}. We found
clear signatures of Bloch-like oscillations for $\alpha > 2$ and their
absence for $\alpha < 2$. The period of the oscillations agrees well with
the period in an ideal Bloch band. Thus, the amplitude of the oscillations
provides a direct way to estimate the energy difference between the two
mobility edges of the delocalized phase. We found that this energy
difference is in perfect agreement with the results reported in
Ref.~\cite{Moura98}. This finding opens the possibility to experimentally
measure the bandwidth of the delocalized phase. Actual technological
advances made it possible to monitor the amplitude of Bloch oscillations in
uniform semiconductor superlattices~\cite{Lyssenko97}. Recently,
intentionally disordered superlattices were used to demonstrate the absence
of localization in short-range correlated disordered
systems~\cite{Bellani99}. Thus, we conjecture that intentionally disordered
superlattices with long-range correlated disorder would allow for a
clearcut validation of the present results. In addition, metallic rings
threaded by magnetic fields linearly rising in time also display Bloch
oscillations~\cite{Arrachea02}, thus opening new experimental ways to test
our predictions.

\acknowledgments

V.~A.~M.\@ acknowledges support through a NATO Fellowship. Work at Madrid
was supported by DGI-MCyT (MAT2000-0734). Work at Brazil was supported by
CNPq and CAPES (Brazilian research agencies) and FAPEAL (Alagoas State
agency).

\end{document}